\documentclass[aps,pra,twocolumn,showpacs,preprintnumbers,amsmath,amssymb,superscriptaddress,10pt]{revtex4-1}

\usepackage{amsmath,amssymb}
\usepackage{bm}
\usepackage[latin1]{inputenc} 
\usepackage{dcolumn}
\usepackage{graphicx}
\usepackage{subfig}
\usepackage{SIunits}
\usepackage{ae}
\usepackage{color} 
\usepackage{tikz}

\newcommand{\matrise}[1]{\begin{bmatrix} #1 \end{bmatrix}}
\newcommand{\diff}[0]{\text{d}}

\newcommand{\im}{\mathrm{Im}\,}

\newcommand{\vek}[1]{\boldsymbol{\mathbf{#1}}}
\newcommand{\vekh}[1]{\hat{{\boldsymbol{\mathbf{#1}}}}}

\newcommand{\be}{\begin{equation}}
\newcommand{\ee}{\end{equation}}
\newcommand{\ba}{\begin{align}}
\newcommand{\ea}{\end{align}}
\newcommand{\e}[1]{\text{e}^{#1}}

\begin{document}

\title{Higher order multipoles in metamaterial homogenization}

\author{Christopher A. Dirdal}
\author{Hans Olaf Hågenvik}
\author{Haakon Aamot Haave}
\affiliation{Department of Electronic Systems, NTNU -- Norwegian University of Science and Technology, NO-7491 Trondheim, Norway}
\author{Johannes Skaar}
\affiliation{Department of Technology Systems, University of Oslo, P.O. Box 70, NO-2027 Kjeller, Norway}
\email{johannes.skaar@its.uio.no}

\date{\today}

\begin{abstract}
The higher order multipoles above the electric quadrupole are commonly neglected in metamaterial homogenization. We show that they nevertheless can be significant when second order spatial dispersive effects, such as the magnetic response, are considered. In this respect, they can be equally important as the magnetization and quadrupole terms, and should not automatically be neglected.
\end{abstract}

\maketitle

\section{Introduction} \label{sec:Intro} 
The structural freedom in metamaterials have spurred renewed interest into homogenization theories. These are theories that allow for the formulation of effective macroscopic Maxwell's equations in structured media from the exact microscopic ones. The macroscopic equations have effective plane wave solutions in materials with complex structures, where dimensions are well below the wavelength. Despite the similarities between conventional and metamaterial homogenization, it has become evident that certain differences need to be taken into consideration \cite{vinogradov1999,silveirinha2007,silveirinha09,Cho2008,Petschulat08,simovski09,Al`u2011a,Al`u2011,Yaghjian2013374}; in particular, the importance of spatial dispersion. In this paper we would like to add another characteristic feature of metamaterial homogenization to the list: That higher order terms in the expansion of macroscopic polarization, above the electric quadrupole, may have physical significance with respect to the magnetic response of the system. Hence, some of the underlying assumptions regarding the non-importance of the electric quadrupole and higher order terms in both classical \cite{agranovich1984crystal, landau_lifshitz_edcm,Russakoff1970, jackson_classical_1999} and more recent \cite{silveirinha2007,silveirinha09,Al`u2011,Yaghjian2013374} treatments on homogenization, should in some cases be reconsidered when applied to metamaterials.

The scattering of a single cell excited by a plane wave has been discussed extensively in the literature. In the long-wavelength limit the electric dipole term generally dominates. The magnetic dipole and electric quadrupole terms may contribute for resonances where the electric dipole term vanishes by symmetry \cite{Cho2008}. Even higher order multipoles can be important in certain cases \cite{Evlyukhin2011}. 

A periodic metamaterial, however, behaves quite different from a single cell, as the neighboring cells are in each other's near field. It is therefore of interest to investigate the importance of the different multipoles for periodic metamaterials. In Sec. \ref{sec:Homog} we review the needed background on the homogenization procedure and multipoles, in addition to the constitutive relations in the Landau-Lifshitz formulation \cite{landau_lifshitz_edcm,vinogradov2002}. In Sec. \ref{sec:ImportanceHigherTerms} we demonstrate that both magnetic dipole + electric quadrupole, and electric octupole + magnetic quadrupole, may be of the same order in $ka$ and of the same order of magnitude. Here $k$ is the wavenumber and $a$ is the cell size. Analytical results and simulations are shown.

Harmonic fields with angular frequency $\omega$ have been assumed, and the $\text{e}^{-i\omega t}$ dependence is suppressed. For simplicity, we will throughout this article consider structures consisting of non-magnetic inclusions. The medium is assumed to be passive (or in thermal equilibrium in the absence of the field under study \cite{landau_lifshitz_edcm}), i.e., we exclude gain media.


\section{Homogenization and multipoles} \label{sec:Homog}
We consider a periodic metamaterial consisting of cubic unit cells of size $a$, and a single spatial Fourier component of the source, $\vek J_\text{ext}=\bar{\vek J}\e{i\vek k\cdot\vek r}$ with constant amplitude $\bar{\vek J}$. The wavevector  $\mathbf k$ is considered as a free parameter, independent of frequency \cite{agranovich1984crystal,silveirinha2007,Al`u2011}. The microscopic fields are Bloch waves of the form
\begin{equation}\label{eq:Bloch}
\vek e(\mathbf{r}) = \vek u_{\vek e}(\mathbf{r}) \e{i\vek k\cdot\vek r},
\end{equation}
where $\vek u_{\vek e}(\mathbf{r})$ has the same periodicity as the metamaterial. The microscopic fields are homogenized according to
\begin{equation} \label{eq:Average}
\vek E \equiv \langle\vek e\rangle \equiv \frac{\e{i\vek k\cdot\vek r}}{V}\int_V \vek e(\mathbf{r}) \text{e}^{-i\mathbf{k \cdot r}} \diff V,
\end{equation}
where the integral is taken over the volume of a unit cell $V$ (see for example \cite{silveirinha09,silveirinha2007,Al`u2011,Yaghjian2013374}). Application of the averaging \eqref{eq:Average} to the microscopic Maxwell equations give macroscopic Maxwell's equations
\begin{subequations}\label{eq:AvMaxwells}
\begin{align}
i \vek k \times \vek E &= i \omega \vek B, \label{eq:AvMaxwellFarad}\\
i\mathbf{k\times \frac{B}{\mu_0}} &= -i \omega \epsilon_0 \mathbf{E} -i\omega \langle \mathbf{p} \rangle + \mathbf{J}_\text{ext}, \label{eq:AveragedAmpGaus} 
\end{align}
\end{subequations} 
having identified $\vek{j} = -i \omega\vek{p}$ and defined macroscopic fields $\vek E=\langle\vek e\rangle$ and $\vek B=\langle\vek b\rangle$. The effective electromagnetic response of the system is contained in the induced current $-i\omega\langle \mathbf{p} \rangle$, which we shall now expand into multipoles \cite{bladel07,Al`u2011}. For sufficiently small $ka$, with the expansion $\exp(-i\vek k\cdot\vek r)\approx 1-i\vek k\cdot\vek r-(\vek k\cdot\vek r)^2/2 + O(k^3)$ we obtain (to the second order in $k$)
\begin{align}
& \langle \mathbf{p} \rangle
= \frac{\text{e}^{i \mathbf{k\cdot r}}}{V}\int_V \mathbf{p} \e{-i\vek k\cdot\vek r}\diff V \label{eq:AvPMul} \\
&= \frac{\text{e}^{i \mathbf{k\cdot r}}}{V} 
\cdot\left(\int_V \mathbf{p}\diff V - i\vek k\cdot\int_V \vek r\mathbf{p}\diff V - \frac{1}{2}\int_V (\vek k\cdot\vek r)^2 \mathbf{p} \diff V \right) \nonumber\\
&\equiv \vek P - \frac{\vek k\times\vek M}{\omega} - i\vek k\cdot\vek Q + \vek R. \label{eq:Multipole}
\end{align}
Here
\begin{subequations}\label{eq:MultipoleVectors}
\begin{align}
\vek P &= \frac{\text{e}^{i \mathbf{k\cdot r}}}{V}\int_V \mathbf{p}\diff V, \\
\vek M &= -\frac{i\omega}{2}\frac{\text{e}^{i \mathbf{k\cdot r}}}{V}\int_V \vek r\times\vek p\diff V, \label{eq:MultipoleVectorsMagn} \\
\vek Q &= \frac{1}{2}\frac{\text{e}^{i \mathbf{k\cdot r}}}{V}\int_V (\vek r\mathbf{p}+\mathbf{p}\vek r)\diff V, \\
\vek R &= -\frac{1}{2}\frac{\text{e}^{i \mathbf{k\cdot r}}}{V}\int_V (\vek k\cdot\vek r)^2\mathbf{p}\diff V, \label{eq:RTerm}
\end{align}
\end{subequations}
and we have decomposed the tensor $\vek r\mathbf{p}$ into its antisymmetric and symmetric parts,
\begin{align}\label{eq:DecomSymAndAntiSym}
\vek k\cdot\vek r\mathbf{p} &= \vek k\cdot(\vek r\mathbf{p}-\mathbf{p}\vek r)/2+\vek k\cdot(\vek r\mathbf{p}+\mathbf{p}\vek r)/2 \nonumber\\
&= -\vek k\times\vek r\times\mathbf{p}/2 +\vek k\cdot(\vek r\mathbf{p}+\mathbf{p}\vek r)/2.
\end{align}
In addition to the polarization vector $\vek P$, magnetization vector $\vek M$, and quadrupole tensor $\vek Q$, we have included an extra term $\vek R$, corresponding to electric octupole and magnetic quadrupole. Apparently, the magnetization and electric quadrupole terms in \eqref{eq:Multipole} seem to be first order in $ka$, while the $\vek R$ term is second order. However, $\vek M$ and $\vek Q$ are themselves dependent on $\vek k$, so the order and magnitude of the terms need to be examined more closely (Sec. \ref{sec:ImportanceHigherTerms}).

In a linear medium, we can express multipole densities \eqref{eq:MultipoleVectors} with constitutive relations
\begin{subequations}\label{eq:2ndOrderkExp}
\begin{align}
P_i &= \epsilon_0\chi_{ij} E_j + \xi_{ikj}k_kE_j + \eta_{iklj}k_kk_lE_j/(\mu_0\omega^2), \label{P2o}\\
M_{m} &= \omega\zeta_{mj}E_j + \nu_{mlj}k_lE_j/(\mu_0\omega), \label{eq:MExp} \\
Q_{ik} &= i\sigma_{ikj}E_j + i\gamma_{iklj}k_lE_j/(\mu_0\omega^2), \label{eq:QuadExp} \\
R_i &= \psi_{iklj}k_k k_l E_j/(\mu_0\omega^2),
\end{align}
\end{subequations} where summation over repeated indices is implied. In \eqref{eq:2ndOrderkExp} we have included the necessary orders of $k$ such that $\langle \vek p \rangle$ is second order in $k$ upon their insertion in \eqref{eq:Multipole}. For later convenience we have included certain $k$-independent quantities (such as $\mu_0\omega^2$) in the tensor elements. Magneto-electric coupling is taken into account in terms of the tensor elements $\xi_{ikj}$ and $\zeta_{mj}$.  

In the so-called Landau-Lifshitz formulation \cite{landau_lifshitz_edcm}, the response of a linear medium is described by a single, nonlocal, relative permittivity tensor $\epsilon(\omega,\vek k)$, such that 
\begin{equation}\label{eq:NonLocalEps}
 \epsilon_0\vek\epsilon(\omega,\mathbf{k}) \mathbf{E} = \epsilon_0\mathbf{E} + \langle \vek p \rangle.
\end{equation}
Here, \emph{all} terms of $\langle \vek p \rangle$, including those resulting from $\vek M$, $\vek Q$ and $\vek R$, are absorbed into $\vek\epsilon(\omega,\vek k)$. From \eqref{eq:Multipole}, \eqref{eq:2ndOrderkExp} and \eqref{eq:NonLocalEps} we obtain
\begin{align}
&\epsilon_{ij}(\omega,\vek k) - \delta_{ij} = \chi_{ij} + \left(\xi_{ikj} + \sigma_{ikj} - \epsilon_{ikm}\zeta_{mj}\right)k_k/\epsilon_0  \label{eq:nonlocalepscomp} \nonumber \\
&+ \left(\psi_{iklj}+\gamma_{iklj}+\eta_{iklj}-\epsilon_{ikm}\nu_{mlj}\right) k_kk_l c^2/\omega^2, 
\end{align}  
where $\epsilon_{ikm}$ is the Levi-Civita symbol. 

While it may be convenient to have only a single constitutive tensor $\epsilon(\omega,\vek k)$, it is often desirable to express the magnetic response more explicitly by introducing a permeability tensor, related to the second order term in \eqref{eq:nonlocalepscomp} \cite{landau_lifshitz_edcm,silveirinha2007}. Observe that the macroscopic quantities $\mathbf{B}$ and $\mathbf{E}$ are left invariant upon the transformation
\begin{align}\label{eq:Transform}
-i\omega \langle \vek p \rangle \to -i\omega \hat{\vek P} +  i\mathbf{k\times} \hat{\vek M},
\end{align} 
where the new polarization $\hat{\vek P}$ and magnetization $\hat{\vek M}$ are arbitrarily chosen. We can express the left hand side in terms of the non-local tensor $\epsilon(\omega,\mathbf{k})$ by \eqref{eq:NonLocalEps}, and the right hand side in terms of two new tensors $\epsilon$ and $1-\mu^{-1}$, in order to obtain
\begin{align} \label{eq:IntroLocalParam}
\epsilon(\omega,\mathbf{k}) = \epsilon - \frac{c^2}{\omega^2} \vek k \times [1- \mu^{-1}] \times \vek k.
\end{align}
Here, we have used $\hat{\vek M}=\mu_0^{-1}(1-\mu^{-1})\vek B$ and \eqref{eq:AvMaxwellFarad}. If we choose the coordinate system such that $\vek k = k \hat{\vek{x}}$, then \eqref{eq:IntroLocalParam} may be expressed
\be\label{eq:epswo}
\epsilon(\omega,\mathbf{k}) = \epsilon + \frac{k^2c^2}{\omega^2}
\matrise{0 & 0 & 0 \\ 0 & (1- \mu^{-1})_{33} & -(1- \mu^{-1})_{32} \\ 0 & -(1- \mu^{-1})_{23} & (1- \mu^{-1})_{22}}.
\ee
We now assume that the medium has a center of symmetry, such that $\epsilon(\omega,-\mathbf{k})=\epsilon(\omega,\mathbf{k})$ \cite{landau_lifshitz_edcm, agranovich1984crystal}. Thus the odd-order term in \eqref{eq:nonlocalepscomp} vanishes. Comparing \eqref{eq:epswo} with \eqref{eq:nonlocalepscomp} leads to
\begin{align}
1&-\mu^{-1} =  \label{eq:mugammam}\\
    &\matrise{\cdot & \cdot & \cdot \\ \cdot & (\psi + \gamma + \eta)_{3113}- \nu_{213} & -(\psi+\gamma+\eta)_{3112} + \nu_{212} \\ \cdot & -(\psi+\gamma+\eta)_{2113}-\nu_{313} & (\psi+\gamma+\eta)_{2112}+ \nu_{312}}, \nonumber
\end{align}
if we choose to put $\epsilon_{22}=1+ \chi_{22}$, $\epsilon_{33}=1+\chi_{33}$, $\epsilon_{23}=\chi_{23}$, and $\epsilon_{32}=\chi_{32}$. The missing entries in \eqref{eq:mugammam} are a result of the fact that $\vek B$ is transverse, $\vek k\cdot\vek B=0$, and that only the transversal part of $\hat{\vek M}$ contributes to the induced current. Even if there is no center of symmetry, such that the first order term in \eqref{eq:nonlocalepscomp} is present, we obtain \eqref{eq:mugammam} if the first order term is absorbed into $\epsilon$.

In principle the magnetization $\hat{\vek M}$ and associated permeability can be defined in an infinite number of ways, by including any given part of the transversal, induced current. Note, however, that any longitudinal part of the induced current cannot be attributed to the magnetization. In other words, in \eqref{eq:IntroLocalParam}, a $\mathcal O(k^2)$ term must sometimes remain in $\epsilon$. 

The choice in \eqref{eq:mugammam} is somewhat natural, as the magnetization term includes all transversal, induced current, except a part possibly induced by the longitudinal component of the electric field. Eq. \eqref{eq:mugammam} is a generalization of the relation in Ref. \cite{silveirinha2007}. The parameters $\epsilon$ and $\mu$ will be referred to as the Landau-Lifshitz parameters due to their relation to the non-local Landau-Lifshitz permittivity \eqref{eq:NonLocalEps}, and are expressed without any argument in order to distinguish the derived permittivity $\epsilon$ in \eqref{eq:IntroLocalParam} from the non-local parameter $\epsilon(\omega, \vek k)$. 

Note that the magnetization $\vek M$ from \eqref{eq:MultipoleVectorsMagn} and $\hat{\vek M}$ are different; the former expresses the magnetic moment density, while the latter results from the choice \eqref{eq:mugammam}. One can define a permeability from $\vek M$ as well; the difference between such a permeability and the one in \eqref{eq:mugammam} will be due to electric quadrupole, higher order multipoles, and the second order term of the electric polarization.

\section{Importance of higher order multipoles} \label{sec:ImportanceHigherTerms}
The tensors $\nu$, $\gamma$, $\psi$, and $\eta$ relate to $\vek M$, $\vek Q$, $\vek R$, and $\vek P$, respectively, in the manner shown in \eqref{eq:2ndOrderkExp}. As seen in \eqref{eq:nonlocalepscomp} these contribute on an equal footing to the second order effects of $\epsilon(\omega, \vek {k})$ \cite{simovski09}, which may be interpreted as describing the magnetic response of the system according to \eqref{eq:mugammam}. While it is known that the quadrupole tensor $\vek Q$ may be significant \cite{Cho2008,Petschulat08}, we shall now show that $\vek{R}$ too can be physically important. 

\begin{figure}[t]
\center
\subfloat[]{\label{fig:UnitCellBars}\begin{tikzpicture} [scale=3.8]
\fill[black!50!white] (0.15,-0.4) rectangle (0.35,0.4);
\fill[black!50!white] (-0.35,-0.4) rectangle (-0.15,0.4);
\draw[-] (-0.5,-0.5)--(-0.5,0.5)--(0.5,0.5)--(0.5,-0.5)--(-0.5,-0.5);
\fill[black!0!white] (-0.52,-0.48) rectangle (0.52,-0.38);
\node [] at (0,0.3) {$\varepsilon=1$};
\node [] at (-0.25,0.3) {$\varepsilon$};
\node [] at (0.25,0.3) {$\varepsilon$};
\draw[<->] (-0.15,-0.2)--(0.15,-0.2);
\node [below] at (0,-0.2) {$0.3a$};
\draw[<->] (-0.35,-0.44)--(-0.15,-0.44);
\node [left] at (-0.335,-0.43) {$0.2a$};
\draw[<->] (0.35,-0.44)--(0.15,-0.44);
\node [right] at (0.335,-0.43) {$0.2a$};
\draw[->,very thick] (0,0)--(0.6,0);
\node [right] at (0.6,0) {$\vek k$};
\draw[<->] (-0.5,-0.55)--(0.5,-0.55);
\node [] at (0,-0.6) {$a$};
\draw[<->] (0.25,0.4)--(0.25,0.5);
\node [right] at (0.3,0.45) {$g/2$};
\draw[->] (0.8,0)--(0.9,0);
\node [right] at (0.9,0) {$\vekh x$};
\draw[->] (0.8,0)--(0.8,0.1);
\node [right] at (0.8,0.1) {$\vekh y$};
\fill[black!0!white] (-0.79,0) rectangle (-0.8,0.1);
\end{tikzpicture}}

\subfloat[]{\label{fig:splitring}\begin{tikzpicture} [scale=3.8]
\path [draw=none,fill=gray, fill opacity = 1] (0,0) circle (0.45);
\path [draw=none,fill=white, fill opacity = 1] (0,0) circle (0.30);
\draw[-] (-0.5,-0.5)--(-0.5,0.5)--(0.5,0.5)--(0.5,-0.5)--(-0.5,-0.5);
\node [] at (-0.22,0.30) {$\varepsilon$};
\node [] at (-0.37,0.42) {$\varepsilon=1$};
\draw[<->] (-0.5,-0.55)--(0.5,-0.55);
\node [] at (0,-0.6) {$a$};
\draw[->] (0,0)--(-0.30,0);
\node [above] at (-0.14,0) {$0.3a$};
\draw[->] (0,0)--(-0.3182,-0.3182);
\node [right] at (-0.12,-0.15) {$0.45a$};
\draw[->,very thick] (0,0)--(0.6,0);
\node [right] at (0.6,0) {$\vek k$};
\fill[black!0!white] (-0.01,-0.5) rectangle (0.01,-0.29);
\fill[black!0!white] (-0.01,0.29) rectangle (0.01,0.5);
\draw[->] (-0.05,0.27)--(-0.01,0.27);
\draw[->] (0.05,0.27)--(0.01,0.27);
\node [] at (0.02,0.21) {$0.02a$};
\end{tikzpicture}}

\subfloat[]{\label{fig:splitringU}\begin{tikzpicture} [scale=3.8]
\path [draw=none,fill=gray, fill opacity = 1] (0,0) circle (0.45);
\path [draw=none,fill=white, fill opacity = 1] (0,0) circle (0.30);
\draw[-] (-0.5,-0.5)--(-0.5,0.5)--(0.5,0.5)--(0.5,-0.5)--(-0.5,-0.5);
\node [] at (-0.22,0.30) {$\varepsilon$};
\node [] at (-0.37,0.42) {$\varepsilon=1$};
\draw[<->] (-0.5,-0.55)--(0.5,-0.55);
\node [] at (0,-0.6) {$a$};
\draw[->] (0,0)--(-0.30,0);
\node [above] at (-0.13,0) {$0.30a$};
\draw[->] (0,0)--(-0.3182,-0.3182);
\node [right] at (-0.14,-0.16) {$0.45a$};
\fill[black!0!white] (0.1,-0.25) rectangle (0.45,0.25);
\draw[->,very thick] (0,0)--(0.6,0);
\node [right] at (0.6,0) {$\vek k$};
\draw[<->] (0.27,-0.25)--(0.27,0.25);
\node [] at (0.38,0.07) {$0.5a$};
\end{tikzpicture}}

\caption{Different unit cells for the simulations: (a) Two bars; (b) Split ring resonator; (c) C-shaped split ring resonator.}
\label{fig:UnitCells}
\end{figure}
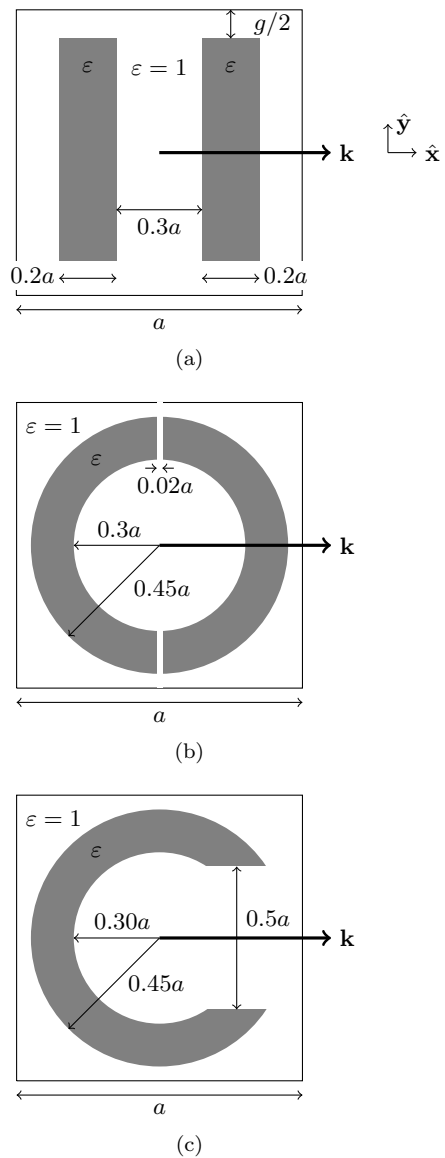

Revisiting the derivation of \eqref{eq:Multipole}, it is tempting to conclude that the magnetization term $-\vek k\times\vek M/\omega$ is first order in $ka$, while $\vek R$ is second order. \emph{However, $\vek M$ is itself dependent on $k$}, being induced by $\vek B=\vek k\times\vek E/\omega$. For  unit cells such as those in Figs. \ref{fig:UnitCells}(a) and \ref{fig:UnitCells}(b), the magnetization $\vek M$ will be zero for $\vek k\to 0$ due to symmetry, provided the origin is located in the middle of the cell. Therefore, $\vek M$ cannot contain any zeroth order term, and must be first order in $ka$. Then the magnetization term $-\vek k\times\vek M/\omega$, quadrupole term, and $\vek R$ term are all second order in $ka$. Even for asymmetric unit cells, such as that in Fig. \ref{fig:splitringU}, the $\vek R$ term can be important when compared to the \emph{second order part} of $-\vek k\times\vek M/\omega$, which is relevant for the magnetic permeability.

We will now demonstrate examples of metamaterial structures where the relevant tensor elements of $\nu$, $\gamma$, $\psi$,  and $\eta$ are of the same order of magnitude. Let the microscopic, relative permittivity of a unit cell be denoted by $\varepsilon(\vek r)$. We first consider a special case which can be treated analytically. For small microscopic susceptibilities $\varepsilon(\vek r)-1$, the field will be almost unperturbed by the cell. Then the microscopic electric field can be approximated by
\begin{align}\label{eq:PlaneWaveSol}
    \vek{e}(\vek{r})= \bar{E}\text{e}^{i\vek{k\cdot r}}\vek{\hat{y}}.
\end{align} 
Taking $\vek k=k\vekh x$, the following relationship may then be observed from \eqref{eq:MultipoleVectors}:
\begin{align}\label{eq:DiffKRel}
    \frac{R_2}{k^2 E_2} = i\frac{\partial}{\partial k} \bigg \{ \frac{Q_{21}}{E_2} \bigg \}= -\frac{\partial}{\partial k} \bigg \{ \frac{M_3}{\omega E_2} \bigg \}= \frac{\partial^2}{\partial k^2}\bigg \{ \frac{P_2}{2E_2} \bigg \},
\end{align} which gives
\begin{align}\label{eq:PsiGammaEta}
    \psi_{2112}= -\gamma_{2112} = -\nu_{312} = \eta_{2112}
\end{align} when compared with \eqref{eq:2ndOrderkExp}. Thus the tensor elements $\psi_{2112}$, $\gamma_{2112}$, $\nu_{312}$, and $\eta_{2112}$ are of the same magnitude in this case.

\begin{figure}[t]
\center
\includegraphics[width=0.31\textwidth]{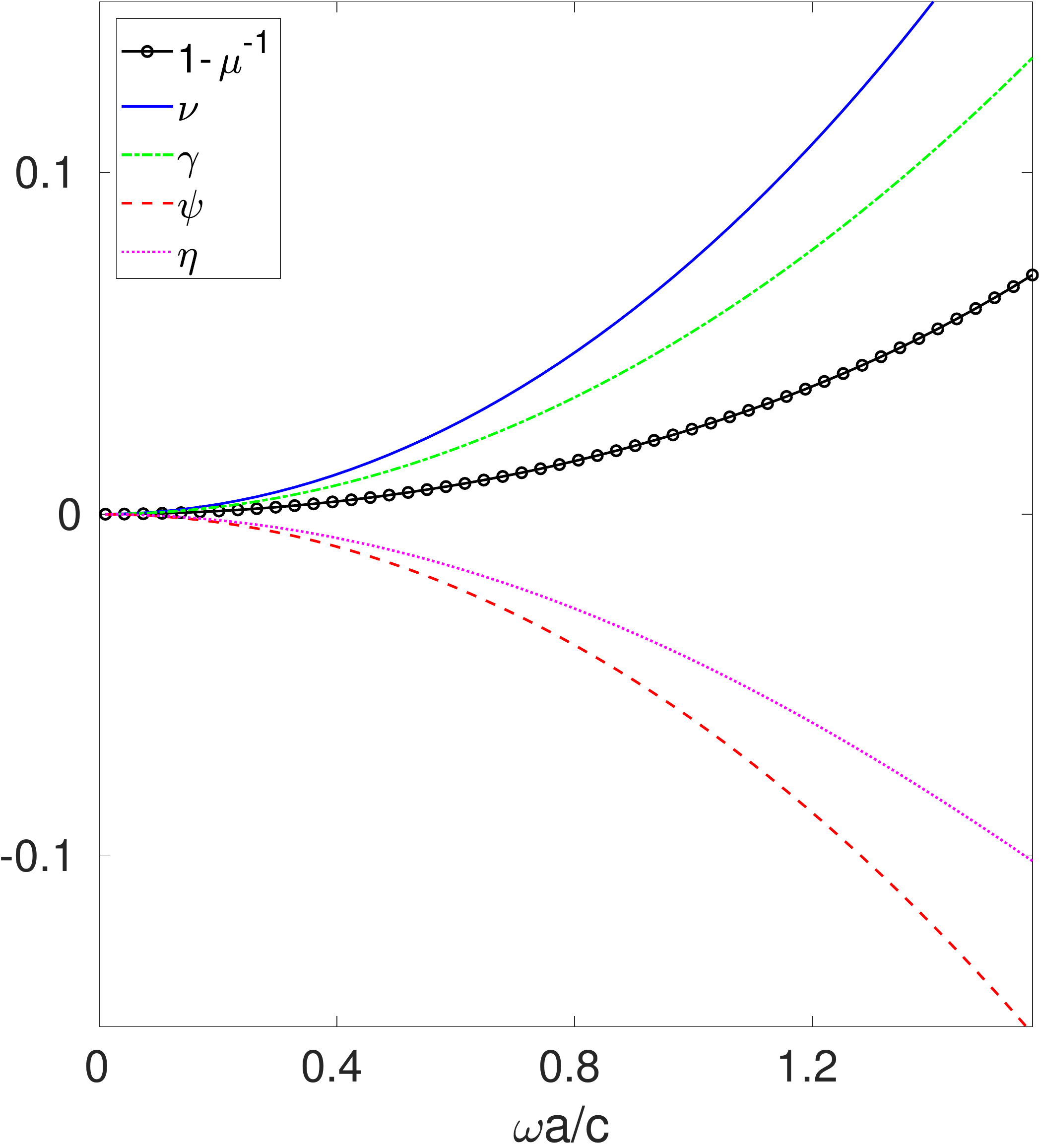}
\caption{The constitutive parameters of the two-bar metamaterial with unit cell as in Fig. \ref{fig:UnitCellBars}, $\varepsilon=16$ and $g=0.2a$.}
\label{fig:ex1}
\end{figure}

\begin{figure}[t]
\center
\includegraphics[width=0.31\textwidth]{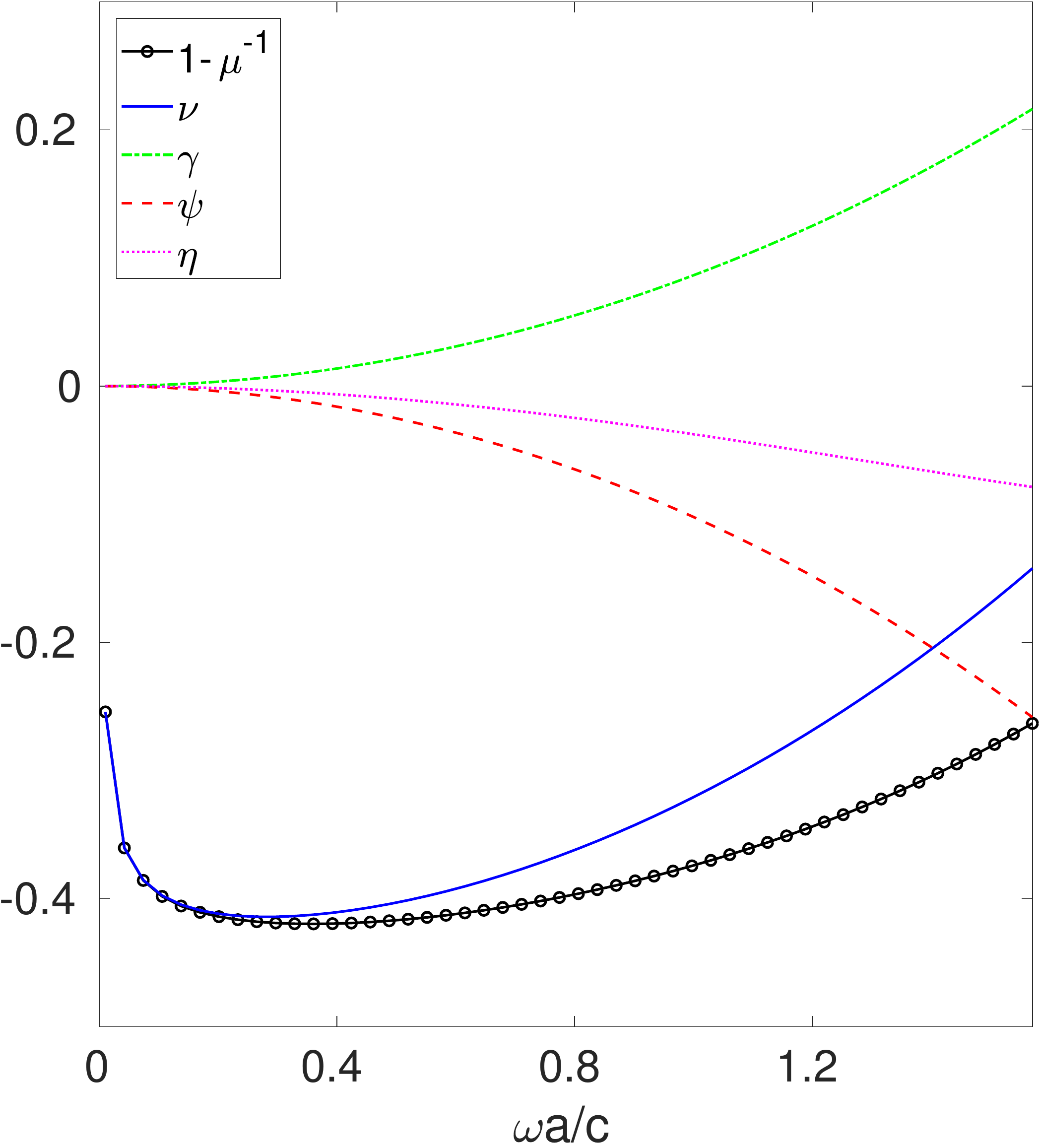}
\caption{Same as Fig. \ref{fig:ex1}, but $\varepsilon$ given by a Drude-Lorentz model of Ag, and $a=10\,\mu$m. Only the real parts are shown; the imaginary parts are $\lesssim 0.1$ times the real parts.}
\label{fig:ex2}
\end{figure}

\begin{figure}[t]
\center
 \subfloat[]{\label{fig:ex3a}
\includegraphics[width=0.4\textwidth]{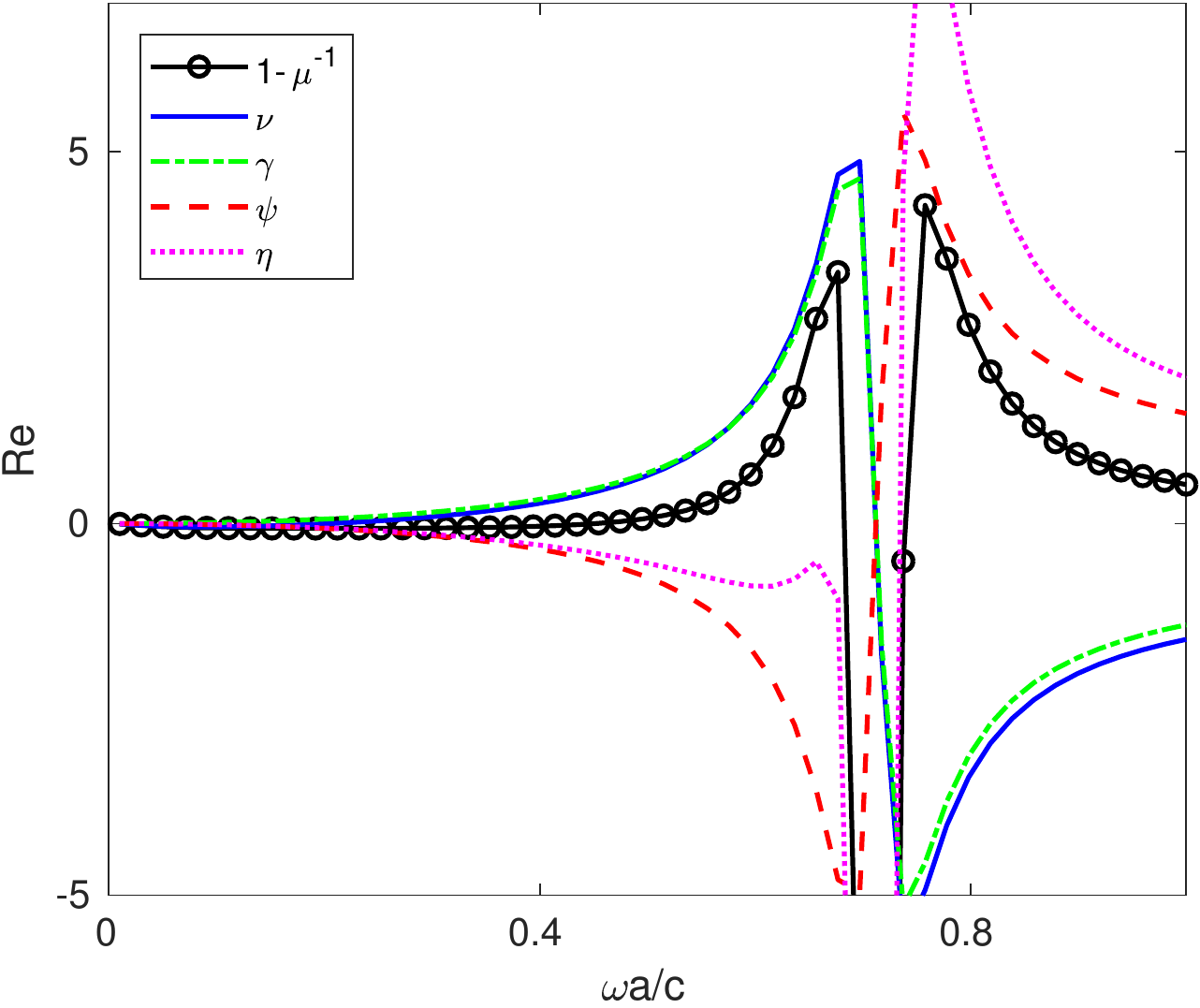}} \\
 \subfloat[]{\label{fig:ex3b}
\includegraphics[width=0.4\textwidth]{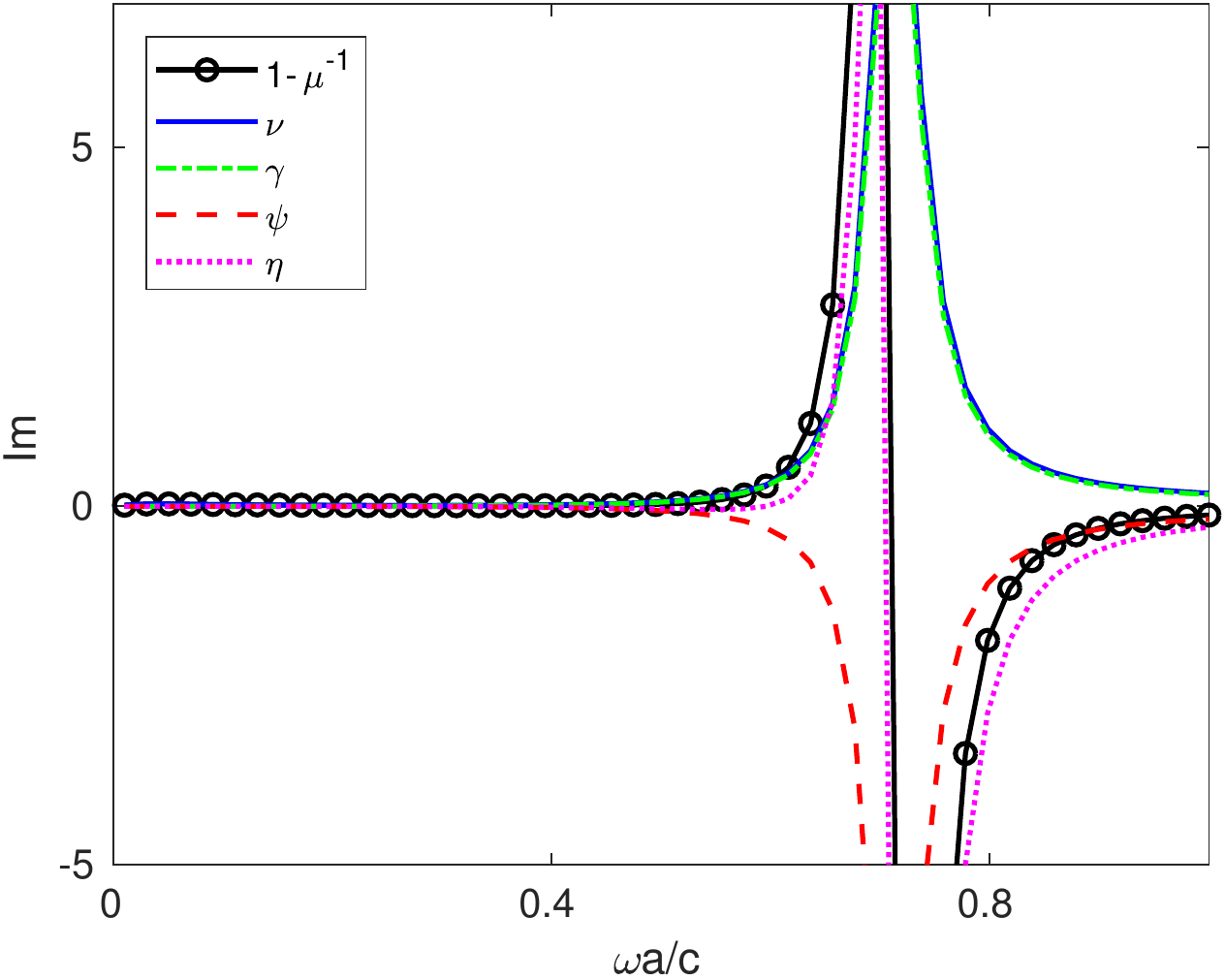}}
\caption{Same as Fig. \ref{fig:ex2}, but $g=0.01a$ and $a=0.2\,\mu$m. Real (a) and  imaginary (b) parts. See the main text for details.}
\label{fig:ex3}
\end{figure}

We now lift the assumption of small microscopic susceptibility, and consider 2d metamaterials with unit cells displayed in Fig. \ref{fig:UnitCells}. A Finite Difference Frequency Domain (FDFD) method is well suited for the problem of computing the microscopic fields, using Bloch-periodic boundary conditions and a source $\vek J_\text{ext}=\bar{\vek J}\exp(ikx)$. The grid is quadratic with $200\times 200$ points. Once the microscopic electric field $\vek e(\vek r)$ and microscopic polarization $\vek p(\vek r)=\epsilon_0\left(\varepsilon(\vek r)-1\right)\vek e(\vek r)$ have been found, we proceed to calculate the multipoles \eqref{eq:MultipoleVectors}. However, solving for the multiple unknowns in \eqref{eq:2ndOrderkExp} generally requires multiple equations. We therefore calculate $\vek{E}$, $\vek{P}$, $\vek M$, $\vek{Q}$, and $\vek{R}$ for two choices of $\bar{\vek{J}}_\text{ext}$, along $\vekh x$ and $\vekh y$, respectively. In order to extract the coefficients in \eqref{eq:2ndOrderkExp}, the field quantities $\vek{E}$, $\vek{P}$, $\vek M$, $\vek{Q}$, and $\vek{R}$ are calculated for three values of $k$ so that first and second order derivatives wrt. $k$ can be obtained. The resulting tensor elements are Taylor coefficients around $k=0$. We are interested in the contributions to $(1-\mu^{-1})_{33}$ from the different multipoles, which according to \eqref{eq:mugammam} is:
\be\label{mucontr}
(1-\mu^{-1})_{33} = \psi_{2112}+\gamma_{2112}+\eta_{2112} + \nu_{312}
\ee
Below, and in the plots, these relevant tensor elements $\psi_{2112}$, $\gamma_{2112}$, $\eta_{2112}$, and $\nu_{312}$ will be denoted $\psi$, $\gamma$, $\eta$, and $\nu$, respectively.

\begin{figure}[t]
\center
\includegraphics[width=0.31\textwidth]{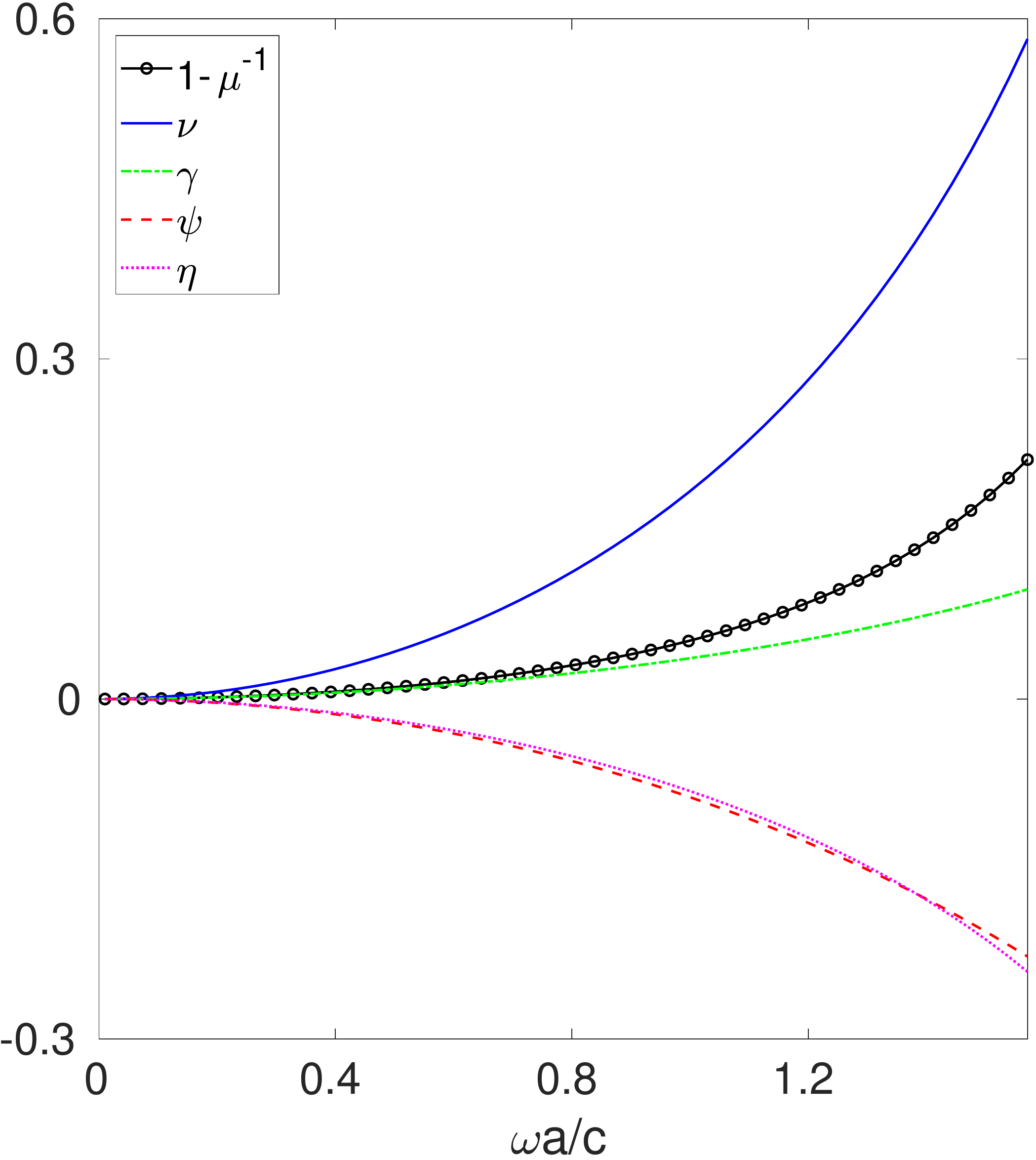}
\caption{The constitutive parameters of the split-ring metamaterial with unit cell as in Fig. \ref{fig:splitring}, $\varepsilon=16$.}
\label{fig:ex4}
\end{figure}

\begin{figure}[t]
\center
 \subfloat[]{\label{fig:ex5a}
\includegraphics[width=0.4\textwidth]{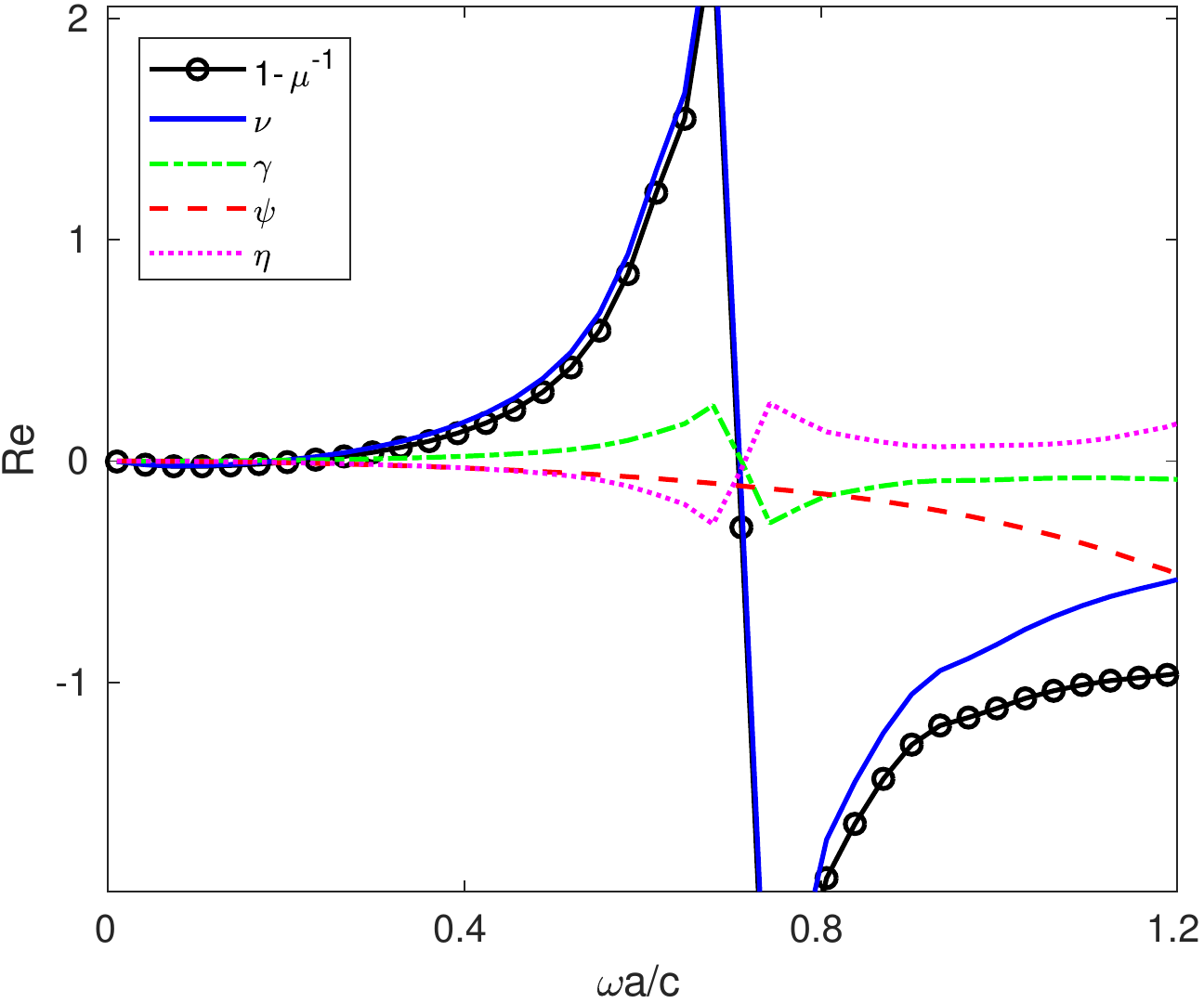}} \\
 \subfloat[]{\label{fig:ex5b}
\includegraphics[width=0.4\textwidth]{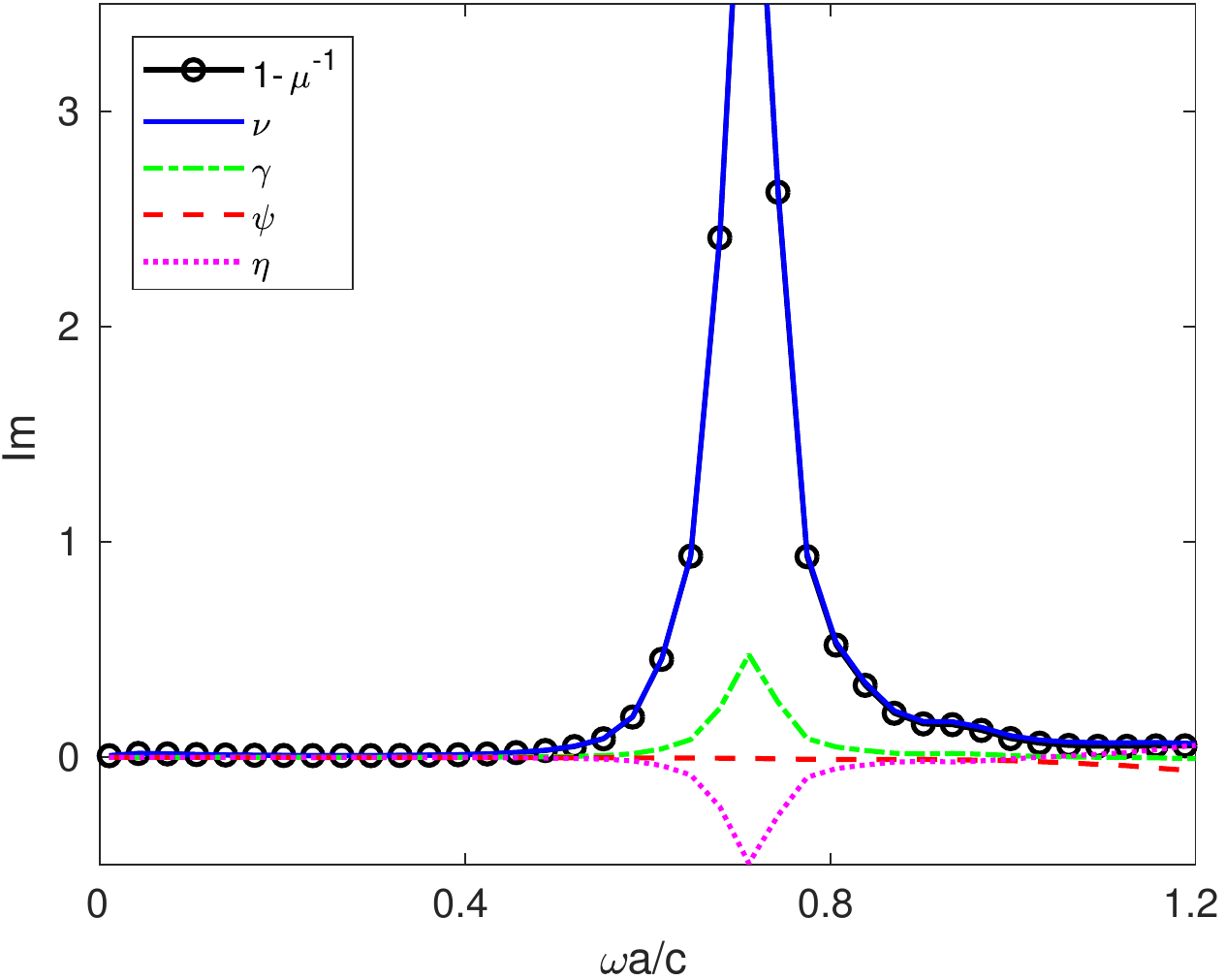}}
\caption{Same as Fig. \ref{fig:ex4}, but the split ring resonator is made from silver, and $a=0.2\,\mu$m. Real (a) and  imaginary (b) parts.}
\label{fig:ex5}
\end{figure}

\begin{figure}[t]
\center
 \subfloat[]{\label{fig:ex6a}
\includegraphics[width=0.4\textwidth]{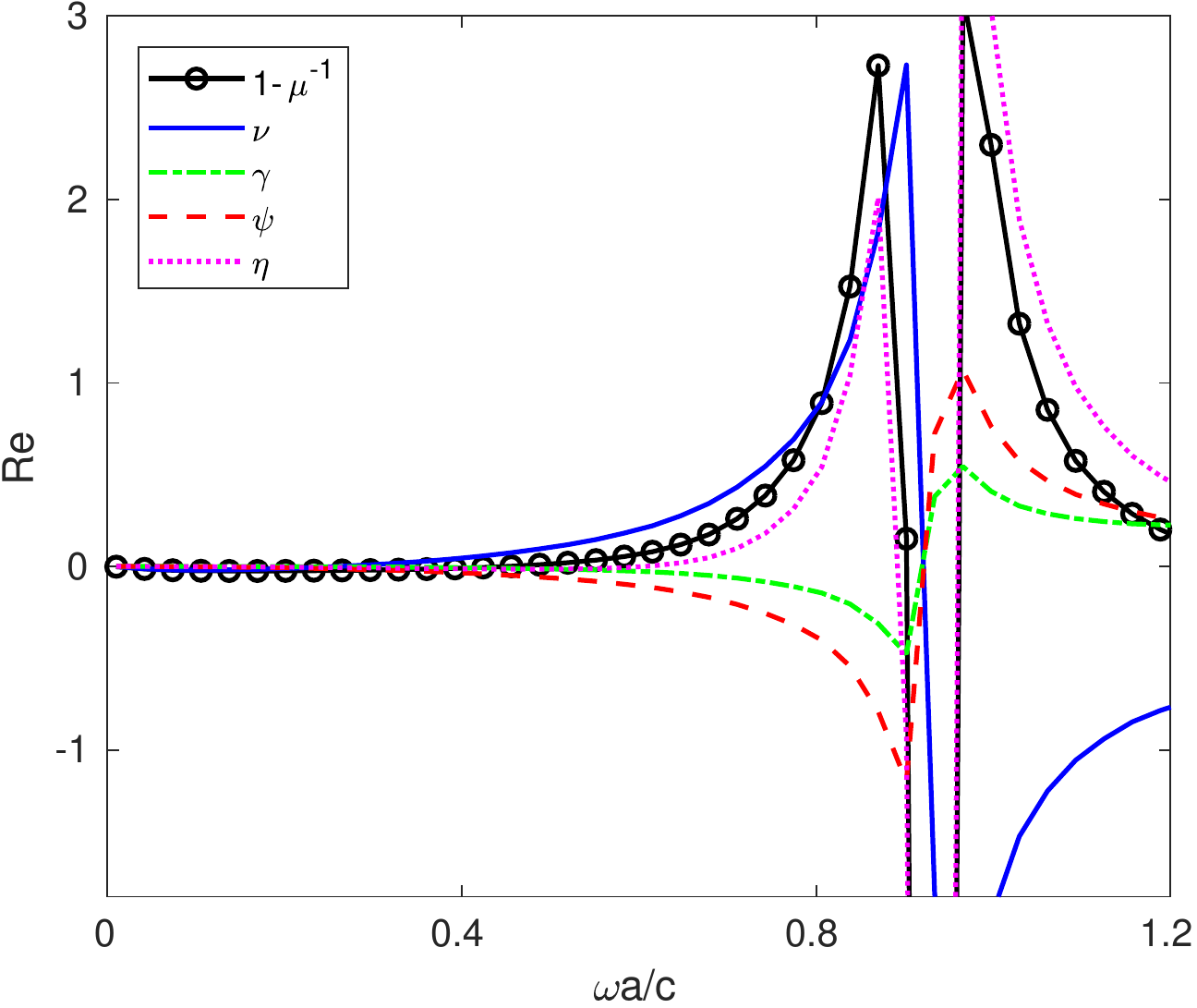}} \\
 \subfloat[]{\label{fig:ex6b}
\includegraphics[width=0.4\textwidth]{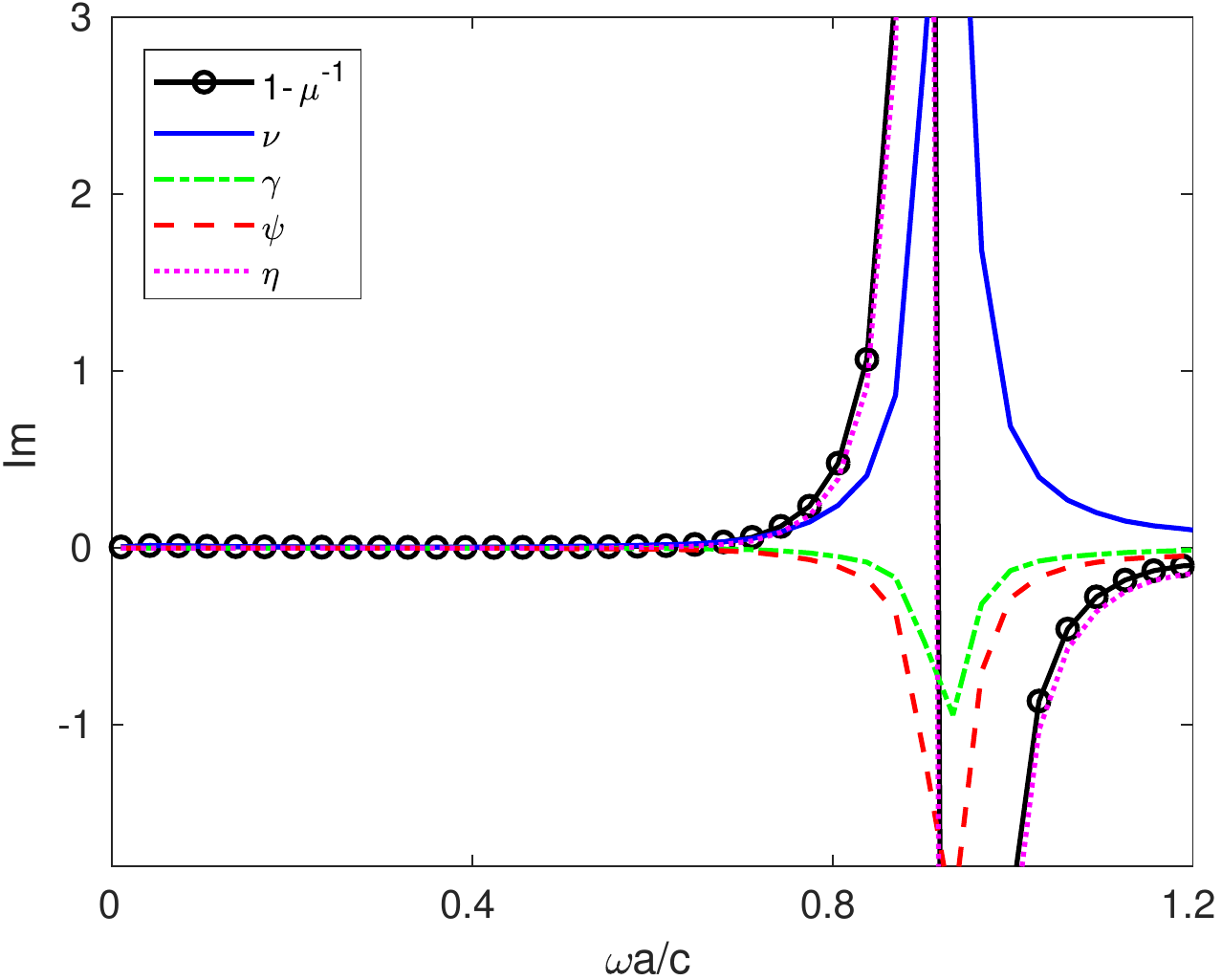}}
\caption{The constitutive parameters of the metamaterial with C-shaped silver split ring resonators, $a=0.2\,\mu$m. Real (a) and imaginary (b) parts. See the main text for details.}
\label{fig:ex6}
\end{figure}

Consider first a metamaterial consisting of the unit cells in Fig. \ref{fig:UnitCellBars}, with $\varepsilon=16$ and $g=0.2a$. The resulting tensor elements are shown in Fig. \ref{fig:ex1}. We observe that $\psi$, $\gamma$, $\eta$, and $\nu$ are of the same order of magnitude. In particular, $|\psi|$ (which results from the higher order multipole term $\vek R$) is approximately equal to $\nu$ (which results from $\vek M$). The sum of the four tensor elements is according to \eqref{mucontr} equal to $1-\mu^{-1}$, which in this case is relatively small.

Next we consider the same system, but let the bars be metallic (Ag), described by a Drude-Lorentz model with parameters from Ref. \cite{rakic98}. The lattice constant is taken to be $a=10\,\mu$m. The resulting tensor elements are shown in Fig. \ref{fig:ex2}. Now the magnetic response is larger. In particular, for small frequencies there is a diamagnetic response,  and $\nu$ dominates ($1-\mu^{-1}\approx\nu$). Thus in this case we can safely ignore the higher order multipoles for small $\omega a/c$.

In Fig. \ref{fig:ex3} we consider the silver bars again, but this time $g=0.01a$, and $a=0.2\,\mu$m. This leads to a resonance. All multipole terms in \eqref{mucontr} contribute substantially, already at relatively small $\omega a/c$. For large $\omega a/c$ we note that $\im\mu<0$, which may seem to violate passivity. However, for spatially dispersive media, the fundamental requirement for passivity is that the total Landau--Lifshitz permittivity has positive imaginary part, $\im\epsilon(\omega,\vek k)>0$ \cite{pitaevskii12}. For our situation, $\epsilon(\omega,\vek k)$ is given by \eqref{eq:epswo}, and the relevant element is
\be\label{epswk22}
\epsilon_{22}(\omega,\vek k) = \epsilon_{22} + \frac{k^2c^2}{\omega^2}\left(1-\mu^{-1}\right)_{33}.
\ee
It can be verified numerically that $\im\epsilon_{22}>0$ for $ka<1$. For large $ka$, the right-hand side of \eqref{epswk22} can be negative, which means that the result is unphysical: For large $ka$, the $\mathcal O((ka)^3)$ terms and higher, which are ignored in the expansion \eqref{eq:Multipole}, will be significant, and will restore a positive value of $\im\epsilon(\omega,\vek k)$. In other words, the results for large frequencies in Fig. \ref{fig:ex3} are only valid for small $ka$. This is a region which is only possible to attain with a suitable set of sources, and is of limited physical relevance.

In Fig. \ref{fig:ex4} we have considered a dielectric split ring structure, with $\varepsilon=16$ (Fig. \ref{fig:splitring}). As for the dielectric bars, the multipole constitutive parameter $\psi$, and also $\eta$, are of the same order of magnitude as $\nu$. The magnetic response is however weak. For the split ring resonator made of silver (Fig. \ref{fig:ex5}, $a=0.2\,\mu$m), the situation is different. As is well known from earlier literature we have a strong resonance, and the magnetic response as given by $\nu$ dominates.

Finally, we consider a C-shaped silver split-ring resonator metamaterial, with a broken mirror symmetry about the $yz$-plane (consisting of unit cells as in Fig. \ref{fig:splitringU} with $a=0.2\,\mu$m). Now all constitutive parameters are of the same order of magnitude. Similarly to the example in Fig. \ref{fig:ex3} we have a region for frequencies $\omega a/c\sim 1$ where $\im\mu$ is negative, while $\im\epsilon(\omega,\vek k)$ is positive for sufficiently small $ka$. As discussed for Fig. \ref{fig:ex3} above, this means that unless $ka$ is small, the $\mathcal O((ka)^2)$ model in the expansion \eqref{eq:Multipole} is not sufficient. 

Since the metamaterial in Fig. \ref{fig:splitringU} does not have a center of symmetry, there will be magnetoelectric coupling in this medium, as described e.g. by a nonzero $\zeta_{mj}$ in \eqref{eq:MExp}. However, the total effect as measured by the first order term in \eqref{eq:nonlocalepscomp} turns out to be vanishing small compared to the second order term, for $ka$ in the range of simulated frequencies $0.01 \leq \omega a/c \leq 1.2$.

\section{Discussion and Conclusion}\label{sec:Concl}
The magnetic permeability can be seen as a $\mathcal O(k^2)$ term in the Landau--Lifshitz total permittivity $\epsilon(\omega,\vek k)$. Not only the magnetic dipole term, but also the electric quadrupole term and the electric octupole--magnetic quadrupole term contribute to $\epsilon(\omega,\vek k)$ to order $\mathcal O(k^2)$. We demonstrate that this contribution can be of the same order of magnitude as that from the magnetic dipole, which means that these higher order multipoles should not automatically be neglected.

Assuming $\vek k=k\vekh x$ we note that the electric octupole---magnetic quadrupole term $\vek R$ results from the even part of $\vek p(\vek r)$ with respect to $x$, while $\vek M$ and $\vek Q$ terms get their contribution from the odd part. Thus there exist current distributions where $\vek R$ is negligible compared to the $\vek M$ and $\vek Q$ terms. In the absence of such odd symmetry, however, the second order terms in $k$ of the three terms can be of the same order of magnitude.

For the dielectric split-ring structure, the $\vek R$ term is important, since the microscopic current distribution will mainly have an even part. For metals, however, the circulating component of the current will be larger, giving a larger magnetic dipole moment. When the symmetry about the $yz$-plane is disturbed (as will be the case for the C-shaped split ring resonator), the $\vek R$ term will again be important.

The multipole terms $\vek M$, $\vek Q$, and $\vek R$ are dependent on the choice of origin. When the origin is moved, the relative sizes of the terms are altered in such a way that the total Landau-Lifshitz tensor $\epsilon(\omega,\vek k)$ is unaltered. We have let the origin be located in the center of the unit cell.

%


\end{document}